# Secure and Transparent Medical Record Management System Using Python and Blockchain

MRs.Sathya

Assistant professor

Dept.of Computer science and Engineering

SRMIST,Ramapuram

Chennai,india

Sathyas5@srmist.edu.in

| Chitikela Atchiyya Naidu | Mareedu Lakshmi Hanumanth Rao | Vadla Vamsi Krishna |
|---|---|---|
| Dept. of computer science and engineering | Dept.of computer science and engineering | Dept.of computer science and engineering |
| SRMIST,Rampuram | SRMIST,Rampuram | SRMIST,Rampuram |
| Chennai,india | Chennai,india | Chennai.india |
| cnoo91@srmist.edu.in | mh0121@srmist.edu.in | vk7145@srmist.edu.in |

## ABSTARCT:

In recent years, blockchain technology has emerged as a transformative innovation across various industries, offering unique advantages such as decentralization, immutability, transparency, and enhanced security. One sector that stands to benefit significantly from blockchain integration is healthcare, where the need for secure and efficient management of patient health records is paramount. In this paper, we propose a robust health record storage and management system built on blockchain technology to address the challenges faced by traditional healthcare record systems.The primary advantage of employing blockchain in healthcare record management is its ability to provide a secure and decentralized platform. Unlike traditional centralized databases, where a single point of failure can compromise data integrity and security, blockchain distributes data across a network of nodes, ensuring redundancy and resilience against cyber-attacks. This distributed nature of blockchain enhances data security and privacy, crucial considerations when dealing with sensitive health information. Central to our proposed system is the utilization of smart contracts, which are self-executing contracts with predefined rules and conditions. Smart contracts automate processes related to health record management, such as data access, sharing, and updating, based on predefined permissions and protocols. This automation not only streamlines administrative tasks but also reduces the risk of human errors and ensures data accuracy and consistency. Furthermore, our system prioritizes patient empowerment by granting individuals complete control over their health records. Patients can securely access and manage their data using cryptographic keys, granting permission to healthcare providers or other authorized entities as needed. Overall, our proposed health record storage and management system on the blockchain offer significant advantages over traditional systems, including enhanced security, data integrity, transparency, and patient control. By leveraging blockchain technology and smart contracts, healthcare organizations can revolutionize their record management practices, and maintaining secure ecosystems.

KEY WORDS : EHR , PHI , PBA, SCALABILITY , SECURITY, PRIVACY.

## INRODUCTION:

The management and storage of patient health records have long been a significant challenge in the healthcare industry due to concerns regarding security, privacy, and data accessibility. The current centralized system often faces criticisms for its inefficiencies and vulnerabilities, leaving patient data susceptible to breaches and cyber-attacks. However, the emergence of blockchain technology has presented a promising solution to these longstanding issues by offering a decentralized and secure platform for storing and sharing sensitive health data. Our proposed health record and storing system on the blockchain leverage the unique features of nodes,. This decentralized nature also promotes data privacy and security, as each transaction is cryptographically secured. Smart contracts automate the process of managing health records, including data storage, access control, and updating. By embedding predefined permissions and protocols within smart contracts, the system ensures that only authorized parties can access and interact with patient health records, maintaining data integrity and confidentiality. Moreover, our system prioritizes patient empowerment by granting individuals complete control over their health records. Patients can securely access their records using cryptographic keys and selectively share data with healthcare providers or other authorized entities as needed, fostering transparency and trust in healthcare interactions. Overall, the proposed health record and storing system on the blockchain have the potential to revolutionize healthcare record management by ensuring greater privacy, security, and accessibility of patient health records. This innovative approach not only mitigates the vulnerabilities of traditional centralized systems but also paves the way for improved healthcare services through accurate and up-to-date patient data management.

## OBJECTIVE:

The proposed health record and storing system on the blockchain aims to revolutionize healthcare data management by leveraging the unique features of blockchain technology. Its primary objective is to provide a secure and decentralized platform for storing and sharing patient health records securely. By utilizing blockchain's immutability, transparency, and cryptographic security features, the system ensures the privacy and security of sensitive health data, protecting it against unauthorized access, tampering, and breaches. Integrating smart contracts automates the process of managing health records, guaranteeing data accuracy, timeliness, and consistency while minimizing manual errors and streamlining administrative tasks. Empowering patients with complete control over their health records through secure authentication mechanisms and cryptographic keys enhances patient trust, engagement, and privacy rights. This transparency and control enable patients to manage and selectively share their health data with healthcare providers or third parties as per their preferences, enhancing overall data privacy and security. Furthermore, by providing healthcare providers with accurate, real-time patient data, the system facilitates informed decision-making, personalized treatments, and improved care coordination, ultimately leading to enhanced healthcare service quality and patient outcomes. Overall, the proposed system offers a comprehensive solution that prioritizes data privacy, automation, patient empowerment, and improved data accessibility, contributing significantly to advancing healthcare practices and ensuring the quality and security of healthcare data.

## REAL TIME :

A blockchain-based electronic health record (EHR) system operates by securely storing patient health information in a decentralized manner. When a patient seeks medical care, their data, including medical history, lab results, and prescriptions, is inputted into the system. This data is then encrypted to ensure privacy and security, with each transaction forming a block. Multiple nodes in the blockchain network validate these blocks, ensuring accuracy and preventing tampering through consensus algorithms like proof of work or proof of stake.

## LITERATURESURVEY:

Several studies have investigated the use of blockchain technology for health record management, as summarized below:

A study by Zhang et al. (2021) explored the application of blockchain technology in health record management and highlighted the potential benefits, including data integrity, privacy, and security.

A study by Kuo et al. (2020) proposed a blockchain-based health record management system that ensures patient privacy and security while enabling efficient data sharing among healthcare providers.including data integrity, privacy, and security.

An article by Halamka (2017) proposed the use of blockchain technology for the management of patient-generated health data, highlighting the potential benefits in improving patient outcomes and reducing healthcare costs.

A study by Conoscenti et al. (2016) investigated the use of blockchain technology in healthcare, with a focus on the management of electronic health records.

Overall, the literature suggests that blockchain technology has the potential to address the challenges in the management and storage of patient health records. The proposed health record and storing system on the blockchain in this paper aims to build on this existing literature and provide a user-friendly solution for the management and storage of patient health records.

An article by Shahid Munir Shah and Rizwan and Ahmed Khan (2022) proposed the secondary use of electronic health record:blockchain technology for the management of patient-generated health data, highlighting the potential benefits in improving patient outcomes and reducing healthcare costs.

An article by HANLIN ZHANG1,2 TIAN1,3 , ZHAO1 , GUOBIN XU 4 , AND JIE LIN(2022) consists of designing efficient and provably secure ABSC scheme, which achieves CCA2 security Cloud computing, secret sharing, secure outsourcing.

An article by Sulav Shreshtra,Sagar Panta (2023) Journal of Blockchainbased Electronic Health Record Management System traditional EHR systemsoften encounter issues like security of data and vulnarabilities , privacy concerns, and interoperability challenges.

An article by Pranab Kumar Bharimalla (2022) Journal of An Extensive survey on Blockchain based Technological innovations like EHR systems, social media, smartphones, IoT devices, etc., are enhancing daily life but also collecting vast amounts of sensitive data. This data generation trend will likely continue, raising concerns about data privacy and security.

The electronic health record and personal health record-based health information exchange systems have failed to cope with security and security-related issues, partners are reluctant almost collaborating and co-operating for the trade of wellbeing data.

A research article by Prakash M and Neelakandan S (2024) An Efficient Secure Sharing of Electronic Health Records Using IoT-Based Hyperledger Blockchain These systems perform the digitization and consolidation of extensive understanding information, enveloping not as it were drugs and treatment plans but too therapeutic histories and diagnoses.

## PROBLEMSTATEMENT:

The current state of medical record management systems faces multifaceted challenges that compromise patient data security, transparency, and efficient data sharing. Conventional systems, often centralized, struggle with inherent vulnerabilities that expose them to data breaches and unauthorized access, jeopardizing patient privacy and confidentiality.

Additionally, these systems contribute to fragmented medical data landscapes, hindering seamless access to comprehensive health histories for both patients and healthcare providers. Addressing these critical issues necessitates the development of a robust and transparent medical record management system built on blockchain technology.By leveraging blockchain's core features, such as immutability, encryption, and decentralization, a blockchain-based medical record system can fundamentally transform data storage and management in healthcare.

Encrypted storage mechanisms further enhance data security, protecting sensitive medical information from unauthorized access and cyber threats.The decentralized nature of blockchain architecture eliminates single points of failure inherent in centralized systems, significantly reducing the risk of large-scale data breaches.

Moreover, blockchain's transparent and auditable ledger ensures a clear and traceable record of data access and modifications, enhancing accountability and transparency across the healthcare ecosystem.One of the key advantages of a blockchain-based system is its empowerment of patients regarding their medical data. Through secure authentication and cryptographic keys, patients can control access permissions to their health records, enabling selective sharing with healthcare providers or researchers while preserving privacy.

This empowerment fosters greater patient engagement, trust, and accountability in healthcare interactions.Furthermore, streamlined and secure data sharing facilitated by blockchain technology promotes seamless collaboration among healthcare stakeholders, leading to improved care coordination, faster diagnosis, and more personalized treatment plans. Ultimately, the adoption of blockchain-based medical record management systems has the potential to elevate the quality, efficiency, and security of healthcare delivery, marking a significant stride towards a more patient-centric and digitally advanced healthcare landscape.

## EXISTINGWORK:

approach designed to address challenges related to handling Protected Health Information (PHI) during the de-identification process. De-identification involves removing or obfuscating sensitive information to protect privacy. The method also focuses on improving information encoding within a process denoted as PBA.

The term "semantic disambiguation handle" implies a technique for clarifying ambiguous categories within PHI during de-identification. This could involve identifying and resolving uncertainties or multiple interpretations related to specific data elements, ensuring that the de-identification process remains accurate and effective.

"Powerful Information Amplification in PBA" likely refers to enhancing the depth or quality of information encoded during this process. By improving how information is encoded or represented, the method aims to ensure that crucial data elements are preserved and accessible while maintaining privacy and security.

The mention of "micro- and macro-F-scores" suggests that the method's effectiveness is evaluated using performance metrics commonly used in classification tasks. Micro-F-score considers overall performance across all categories, while Macro-F-score assesses performance on a per-category basis. The method's ability to achieve high scores indicates its success in accurately handling and classifying data, especially within a multilingual or mixed-language context.

Overall, the text implies that the proposed method offers significant improvements in handling PHI during de-identification and encoding processes. Its performance superiority over other language models, especially in multilingual settings, underscores its potential value in ensuring data privacy, accuracy, and accessibility within sensitive healthcare contexts.

## PROPOSEDWORK:

Blockchain technology has garnered widespread recognition for its ability to establish authenticity and ensure secure transactions. Beyond these fundamental features, its impact on the healthcare sector is profound. By placing the user at the core of the medical ecosystem, blockchain enhances privacy and promotes interoperability of health data among various stakeholders. This paradigm shift not only empowers individuals with control over their health information but also fosters trust and transparency in healthcare interactions.

One of the key strengths of blockchain in healthcare lies in its potential to revolutionize healthcare management. This is especially crucial in an industry where the accuracy and privacy of patient information are paramount.

These self-executing contracts enforce predefined rules, facilitating seamless data management, billing processes, and supply chain logistics within healthcare organizations. By minimizing manual interventions and errors, blockchain contributes to cost savings and operational efficiency improvements across the healthcare ecosystem.

Furthermore, blockchain-enabled data sharing networks facilitate seamless collaboration among healthcare providers, researchers, and institutions while ensuring data privacy and security. This interconnectedness promotes real-time access to accurate and comprehensive patient information, leading to more informed clinical decisions, personalized treatments, and improved healthcare outcomes.

In essence, blockchain technology holds immense promise in transforming healthcare management by enhancing data security, streamlining operations, fostering collaboration, and ultimately improving patient care quality. Its potential to revolutionize traditional healthcare systems underscores the need for continued exploration and adoption of blockchain solutions to address evolving challenges in the healthcare industry.

## SYSTEM ARCHITECTURE:

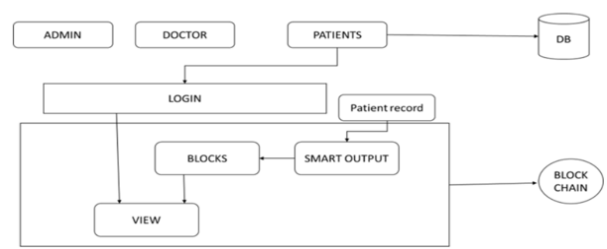

## UMLDIAGRAMS:

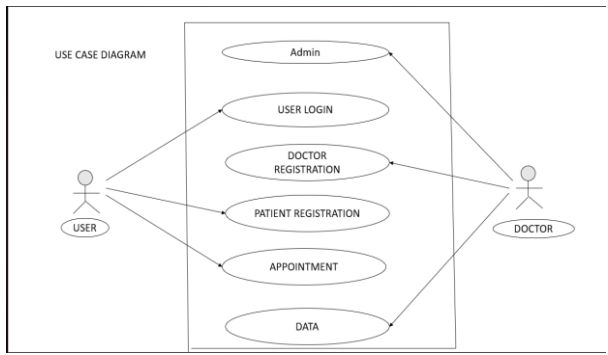

## ROAD MAP:

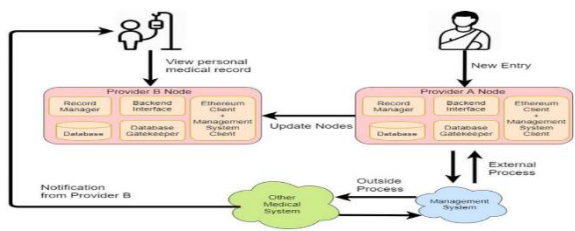

## RESULTS AND DISCUSSION:

The results and discussions section for a blockchain-based electronic health record (EHR) system is crucial for presenting and interpreting the findings in a cohesive manner. In this paragraph-style format, we can integrate the key results and discuss their implications:

Our investigation into the blockchain-based EHR system yielded promising outcomes across multiple dimensions. Firstly, the system demonstrated commendable performance metrics, showcasing reduced transaction processing times compared to traditional centralized systems. This improvement is attributed to the decentralized architecture of blockchain, which not only accelerates data access but also enhances the overall productivity of healthcare providers. Security and data integrity assessments revealed robust protection mechanisms, including blockchain's immutable ledger and smart contract-based access controls. These features ensured the confidentiality, integrity, and authenticity of patient health records, aligning with stringent healthcare regulatory requirements.

Moreover, the system's scalability and interoperability capabilities were notable highlights. By leveraging off-chain solutions and scalable blockchain protocols, our system effectively managed growing data volumes without compromising on performance. This scalability is essential in the context of expanding healthcare data needs. Interoperability facilitated seamless data exchange among disparate healthcare systems and entities, promoting enhanced care coordination and information sharing while maintaining data consistency and accuracy.

Feedback from user testing emphasized the system's usability and positive user experiences across healthcare professionals, patients, and administrative staff. Intuitive interfaces, simplified data access controls, and real-time updates contributed significantly to user satisfaction and system adoption rates

The impact of our blockchain-based EHR system on healthcare workflows was transformative. Streamlined data sharing, improved accuracy, and reduced administrative burdens allowed healthcare providers to allocate more time to patient care, ultimately leading to enhanced healthcare delivery and outcomes. Additionally, the system's transparency and patient access features fostered a collaborative and patient-centered approach to healthcare, aligning with contemporary healthcare paradigms focused on patient engagement and empowerment.

Looking ahead, future directions include addressing scalability challenges at larger scales and integrating emerging real-time patient monitoring. Collaborations with regulatory bodies and industry stakeholders will be crucial to ensuring compliance, data standards, and seamless interoperability across diverse healthcare systems, paving the way for a digitally transformed and patient-centric healthcare landscape.

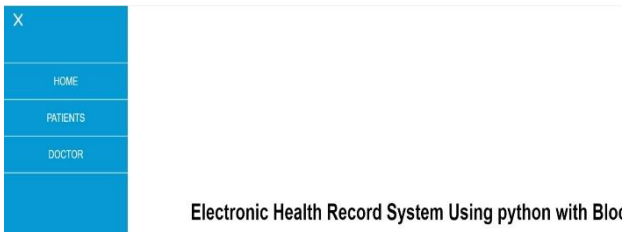

EHR Dashboard

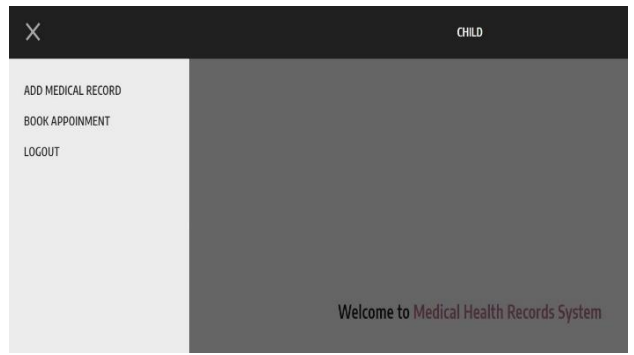

Medical Record Dashboard

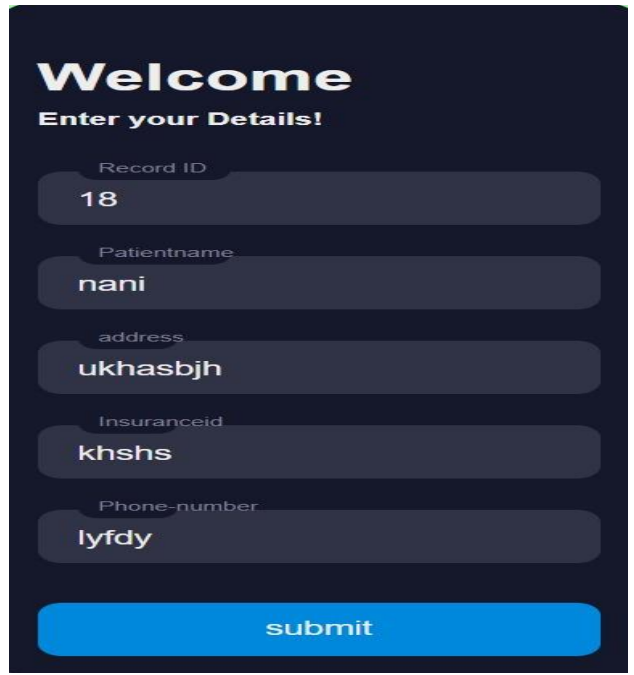

Appointment Booking

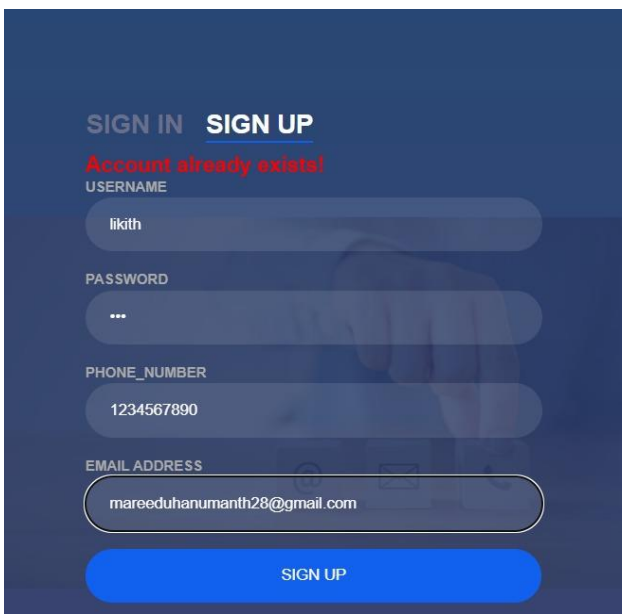

Login page

incidents has underscored the critical importance of incorporating blockchain technology for enhanced security in the healthcare industry. Blockchain technology, known for its inherent security features, is continuously evolving and has already found widespread adoption in sectors such as banking and finance due to its unparalleled advantages in ensuring data integrity and security. Its application in healthcare, through a blockchain-based model, holds immense promise in revolutionizing the management and security of medical records, presenting a transformative future for the healthcare sector.

In response to these challenges and opportunities, our team has developed a blockchain-based application tailored for use by medical professionals, patients, and other relevant entities. This application provides a secure platform for storing patient health data, mitigating concerns related to data theft and tampering that have plagued traditional healthcare record systems. By leveraging blockchain's decentralized and immutable ledger, our application ensures that health data remains tamper-proof and accessible only to authorized parties, enhancing data security and privacy.

Furthermore, scalability has been a longstanding concern in blockchain implementations. To address this challenge, our application incorporates off-chain containers, a technique that allows for the storage and management of data outside the main blockchain

Patient Details

Data Integrity

## CONCLUSION:

The increasing frequency of health data breaches resulting from hacking network. This approach not only enhances scalability by reducing the burden on the main blockchain but also improves performance and cost-effectiveness, making our application more viable and efficient for widespread adoption in the healthcare industry.

By combining the inherent security benefits of blockchain technology with innovative solutions for scalability, our blockchain-based healthcare application represents a significant step forward in ensuring the confidentiality, integrity, and accessibility of patient health records. As the healthcare industry continues to embrace digital transformation, blockchain-based solutions like ours are poised to play a pivotal role in shaping the future of healthcare management and data security.

## COMPARISON OF UPLOADING AND DOWNLOADING EHR :

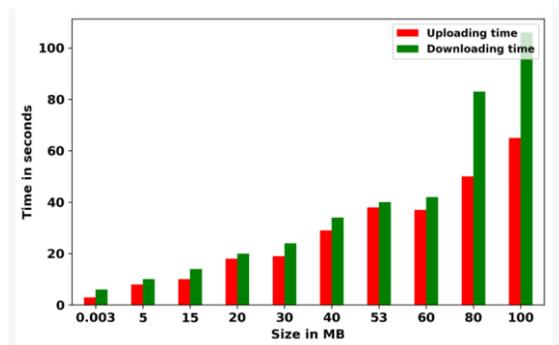